\documentclass[conference]{IEEEtran} 

\usepackage[dvips,dvipdfm,ps2pdf,hyperfigures,colorlinks]{hyperref}
\usepackage{cite}
\usepackage{amsmath,amssymb,graphicx,amsthm}
\usepackage{subfig}

\newtheorem{theorem}{Theorem}[section]
\newtheorem{lemma}[theorem]{Lemma}

\newtheorem{definition}[theorem]{Definition}

\newcommand{\eps}{\epsilon}

\newcommand{\poly}{\operatorname{\text{{\rm poly}}}}

\newcommand{\Fq}{\mathbf{F}_q}

\begin{document}

\title{Optimality of Network Coding in Packet Networks}
\author{%
\IEEEauthorblockN{Bernhard Haeupler}
\IEEEauthorblockA{RLE, CSAIL\\
Massachusetts Institute of Technology\\
Email: haeupler@mit.edu}
\vspace*{-0.5cm}
\and
\IEEEauthorblockN{MinJi Kim}
\IEEEauthorblockA{RLE\\
Massachusetts Institute of Technology\\
Email: minjikim@mit.edu}
\vspace*{-0.5cm}
\and
\IEEEauthorblockN{Muriel M\'edard}
\IEEEauthorblockA{RLE\\
Massachusetts Institute of Technology\\
Email: medard@mit.edu}
\vspace*{-0.5cm}
}

\maketitle

\begin{abstract}
We resolve the question of optimality for a well-studied packetized implementation of random linear network coding, called PNC. In PNC, in contrast to the classical memoryless setting, nodes store received information in memory to later produce coded packets that reflect this information. PNC is known to achieve order optimal stopping times for the many-to-all multicast problem in many settings.

We give a reduction that captures exactly how PNC and other network coding protocols use the memory of the nodes. More precisely, we show that any such protocol implementation induces a transformation which maps an execution of the protocol to an instance of the classical memoryless setting. This allows us to prove that, for \emph{any} (non-adaptive dynamic) network, PNC converges with high probability in \emph{optimal} time. In other words, it stops at \emph{exactly} the first time in which \emph{in hindsight} it was possible to route information from the sources to each receiver individually.

Our technique also applies to variants of PNC in which each node uses only a \emph{finite} buffer. We show that, even in this setting, PNC stops exactly within the time in which in hindsight it was possible to route packets given the memory constraint, i.e., that the memory used at each node never exceeds its buffer size. This shows that PNC, even without any feedback or explicit memory management, allows to keep minimal buffer sizes while maintaining its capacity achieving performance.
\end{abstract}

\section{Introduction}\label{sec:introduction}
It is by now a classical result \cite{li2003linear} that linear network coding archives capacity for multicast and that even choosing a random linear code suffices with high probability \cite{ho2006random,koetter2003algebraic}. The rateless and self-adaptive nature of random linear network coding has been shown particularly beneficial in distributed settings with time-varying network topologies. For these settings, a distributed and packetized network coding (PNC) implementation has been proposed \cite{chou2003practical,lun2008coding} in which nodes keep received packets in memory and forward random linear combinations of these packets whenever they send a packet. The performance of PNC has since been studied in various network and communication models such as: static networks with losses \cite{lun2008coding,lun2005further,Lunthesis,wu2006trellis}, gossip networks \cite{algebraicgossip-deb-med-choute-06-transinf,mosk2006information,vasudevan2009algebraic,borokhovich2010tight} or adversarial dynamic networks \cite{haeupler2010analyzing,petar}. These works feature new and interesting techniques such as connections to queuing networks and Jackson's Theorem \cite{lun2008coding,borokhovich2010tight}, or other stochastic modeling\cite{algebraicgossip-deb-med-choute-06-transinf,petar} and prove (asymptotic) order optimal stopping times that are tight in worst-case examples.  

We show that the optimality of PNC can be understood via a reduction to the classical memoryless setting \cite{ho2006random,koetter2003algebraic}. We prove that in \emph{any} network model, whether it is static, stochastic or non-adaptively adversarial, PNC converges with high probability in \emph{optimal} time: With high probability PNC delivers all information to all receivers at \emph{exactly} the first time-step in which \emph{in hindsight} it was possible to route this information from the sources to each receiver individually. 

Our reduction shows that, for any network coding protocol, it is possible to describe a transformation that captures exactly how the memory of the nodes is used. The transformation, which is induced by a concrete protocol implementation, maps an execution of the protocol to an instance of the classical memoryless setting. 
 This technique also applies to variants of PNC~\cite{haeupler2011onepacket,lun2006analysis} in which each node only keeps a finite amount of packets in active memory. We show that, even in this setting, PNC stops exactly within the time in which in hindsight it was possible to route packets given the memory constraint, i.e., that the buffer at each node never exceeds its active memory size. This shows that PNC, even without any feedback or explicit memory management~\cite{sundararajan2008arq}, optimally uses the limited buffers. 

This paper is organized as follows. We provide a short review of the memoryless network coding results and PNC in Section \ref{sec:nc}. In Section \ref{sec:model}, we introduce our network model. In Section \ref{sec:results}, we present our method to transform a protocol execution into a circuit that captures exactly how a given protocol implementation uses memory. Using this, we prove that PNC and several of its variants are optimal in Section \ref{sec:simulateandoptimality}. Finally, we summarize our contributions in Section \ref{sec:conclusions}.

\section{Network Coding Review}\label{sec:nc}


\vspace*{0cm}\subsection{Memoryless Network Coding Setting} \label{sec:classic}

In the memoryless network coding setting \cite{li2003linear,ho2006random,koetter2003algebraic}, a directed acyclic circuit processes messages from a finite field $\Fq$ (or alternatively $\Fq^l$). A circuit is a directed acyclic hypergraph $\mathcal{C} = (V, A)$. For each node $v \in V$, we denote $\Gamma^+(v)$ as the incoming hyperedges, and $\Gamma^-(v)$ as the outgoing hyperedges. For each $e \in \Gamma^-(v)$, $v$ contains a coding vector $c_e \in \Fq^{\Gamma^+(v)}$. We assume that there is only one node with exclusively outgoing hyperedges, the source node $s \in V$. Assuming an assignment of a message $val(e) \in \Fq$ to each hyperedge $e\in \Gamma^-(s)$, the circuit $\mathcal{C}$ processes information as follows. Each hyperedge can inductively be assigned a message in $\Fq$ by using the rule that the vector associated with an outgoing hyperedge $e$ of $v$ is $c_e \cdot val(\Gamma^+(v))$. In this way, $\mathcal{C}$ defines a linear transform $T(\Gamma^-(s),E') \in \Fq^{\Gamma^-(s) \times E'}$ between the messages $val(\Gamma^-(s))$ and the messages assigned to any subset of hyperedges $E' \subseteq E$.

Reference \cite{li2003linear} shows that if the field size $q$ is large enough, one can choose the $c_e$ such that the rank of $T(\Gamma^-(s),E')$ is equal to the min-cut between $s$ and $E'$ in $\mathcal{C}$. In such a case, any node $v$ with a min-cut of at least $|\Gamma^-(s)|$ can invert $T(\Gamma^-(s),\Gamma^+(v))$ and decode all messages $val(\Gamma^-(s))$. Furthermore \cite{ho2006random,koetter2003algebraic} show that, with high probability, this remains true even if the coding coefficients are chosen uniformly at random. These are the classical results on (random linear) network coding that started this line of research.  

Note that, in this model, timing is irrelevant and each node processes each message only once. References \cite{li2003linear, koetter2003algebraic} show that this setting can be extended to non-acyclic circuits with delays. Nonetheless, nodes remain stateless and memoryless, which is why we refer to these networks as circuits. 

\vspace*{0cm}\subsection{PNC: Distributed Packetized Network Coding}\label{sec:pnc}

In this section, we introduce the PNC protocol~\cite{chou2003practical,lun2008coding} in which, in contrast to the memoryless setting, nodes store received information in memory to later produce coded packets that reflect this information.

Assume that there are $k$ messages from $\Fq^l$ distributed to the nodes. If the PNC protocol is used in a network, any node $u$ communicates by sending packets that contain vectors from $\Fq^{k+l}$ and maintains a subset $S_u \subset \Fq^{k+l}$ of received packets. Initially, $S_u$ is empty for all nodes $u$. When node $u$ initially knows the $i^{th}$ message $s_i \in \Fq^l$, it adds the vector $(e_i,s_i)$ to $S_u$, where $e_i$ is the $i^{th}$ unit vector in $\Fq^k$. If node $u$ is requested to send a packet it sends a random vector from the span of $S_u$. Note that this description is completely independent of any assumption on the network. 

If enough communication takes place among nodes for the system to ``mix'', then for each node $u$ the subspace spanned by $S_u$ will converge to the $k$ dimensional subspace of $\Fq^{k+l}$ given by the $k$ input vectors. Each node can then use Gaussian elimination to recover the input messages.

References \cite{algebraicgossip-deb-med-choute-06-transinf,mosk2006information,vasudevan2009algebraic,borokhovich2010tight,haeupler2010analyzing} provide upper bounds on how quickly this ``mixing'' happens for specific (stochastic) communication models. In this work, we prove a stronger statement that the mixing happens with high probability in optimal time for \emph{any} communication history.

\section{Network Model: Time Expanded Hypergraphs}\label{sec:model}

We consider discrete or continuous time dynamic network topologies where communication links are established synchronously and/or asynchronously. Nodes can potentially send data at different and highly non-regular rates. Links are assumed to have varying delays. We also incorporate broadcast constraints that arise in wireless settings. Our model applies to any static or stochastic model, including arbitrary stochastic link failures, and to adversarial worst-case communication schedules chosen by an oblivious adversary. All these models specify a (distribution over) communication schedules that is independent from the randomness in the coding coefficients. We shall prove a point-wise optimality, i.e., for any instance of a communication schedule, PNC achieves optimal performance. Therefore, throughout the rest of the paper, we assume that there is a specific given communication schedule on which we have to give an optimality proof.

Each communication schedule can be specified as a sequence of \emph{events}, where a node sends or receives packets. We assume that, at each time, a node either transmits or receives a packet. We capture these events using the following definition of a time expanded communication hypergraph. This notion of time expanded hypergraph has been previously used under different names, e.g. continuous trellis \cite{wu2006trellis} or adversarial schedule\footnote{Indeed, Theorem 3.9. in \cite{petar} states a result similar to our main theorem for PNC but with a weaker bound. Instead of proving PNC to be exactly capacity achieving with failure probability $\eps = 1/poly(n)$, their result requires at least $p \cdot l \cdot (\log k + \log \eps^{-1})$ additional capacity. In general, $p$ and $l$ can be of the order of $k$ or even larger making this bound quite loose.} \cite{petar}.

\begin{definition}[Time Expanded Hypergraph]
Consider a network with $n$ nodes, and denote this set of nodes as $V$. A communication schedule from time $0$ to $t$ among nodes in $V$ is captured by the following time expanded hypergraph $G = (V,V',A)$. Let $v\in V$ be a node in the network. We create a vertex copy $v_{t'} \in V'$ for every time $t'\in [0,t]$ when the node $v$ receives or sends at least one packet. If $v$ is transmitting at time $t'$ to nodes $u^1, u^2, ..., u^b$ with associated delay $\Delta_1, \Delta_2, ..., \Delta_b$ respectively, we create a single hyperedge $(v_{t'}, \{u^1_{t'+\Delta_1}, u^2_{t'+\Delta_2}, ..., u^b_{t'+\Delta_b}\}) \in A$.
\end{definition}

Given a network, we consider the following (distributed) many-to-many multicast problem. Messages are generated at nodes in the network. A message can be generated at (multiple) different times at multiple nodes. The goal is to disseminate all the messages to all nodes (or a subset of \emph{destination nodes} $D\subseteq V$) as fast as possible. One example of an application of this problem could be a source distributing a large file (which is divided into small parts) to many receivers. Another application is in sensor networks, where each sensor transmits its measurements at different times.

To formalize this problem, we assume that there are exactly $k$ messages that are vectors of $\Fq^l$. We assume that the nodes employ the PNC protocol of Section \ref{sec:pnc}. Note that this requires each message to have a unique identifier that is known to every node at which the messages is generated. We incorporate the message generation in our network model using the following additional definition. 

\begin{definition}
Let $G = (V,V',A)$ be a communication schedule of a network in which $k$ messages $m_1, \ldots, m_k \in \Fq^l$ are generated. We alter $G$ by adding a supersource node $s$ to $V'$. Furthermore for each message $m_i$ that is generated by nodes $u^1, u^2, ... $ at time $t_1, t_2, ...$ we add a hyperedge $(s, \{u^1_{t_1}, u^2_{t_2}, ...\})$ to $A$. 
\end{definition}

\begin{figure*}
\begin{center}
\subfloat[Network transactions over time]{\label{fig:example_network}\includegraphics[width=0.33\textwidth]{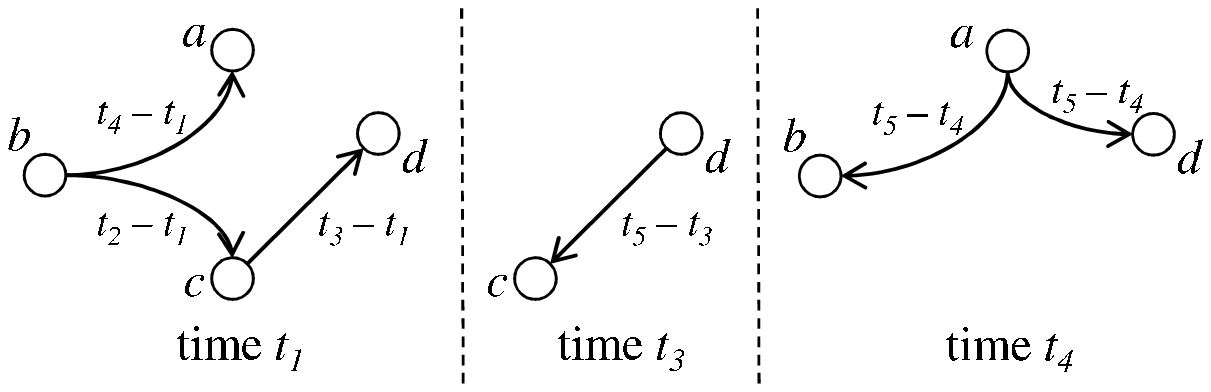}}\hspace*{.1cm}
\subfloat[Time expanded hypergraph]{\label{fig:example_history}\includegraphics[width=0.21\textwidth]{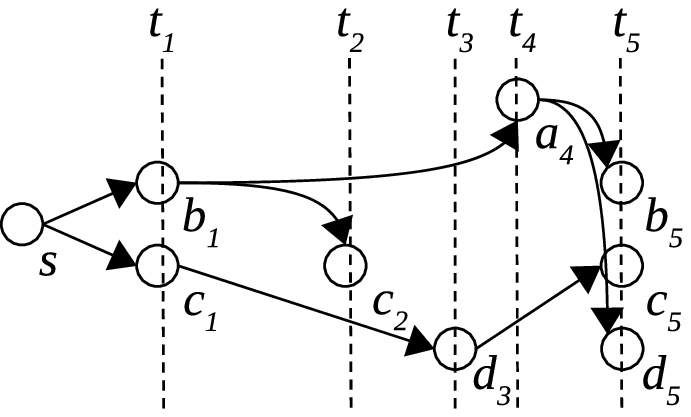}}\hspace*{.1cm}
\subfloat[Information flow hypergraph]{\label{fig:example_infinite}\includegraphics[width=0.21\textwidth]{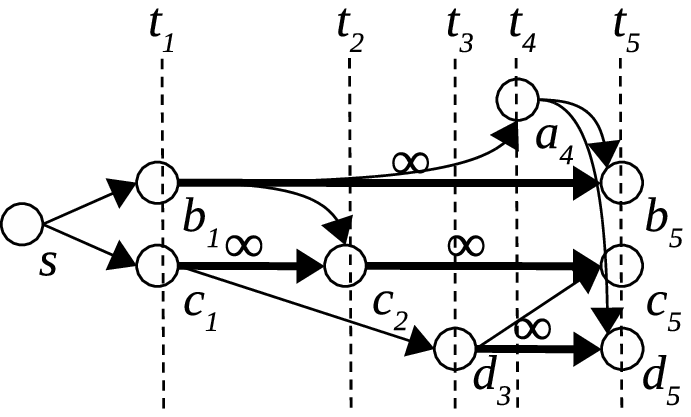}}\hspace*{.1cm}
\subfloat[PNC Transform]{\label{fig:example_fullmemory}\includegraphics[width=0.21\textwidth]{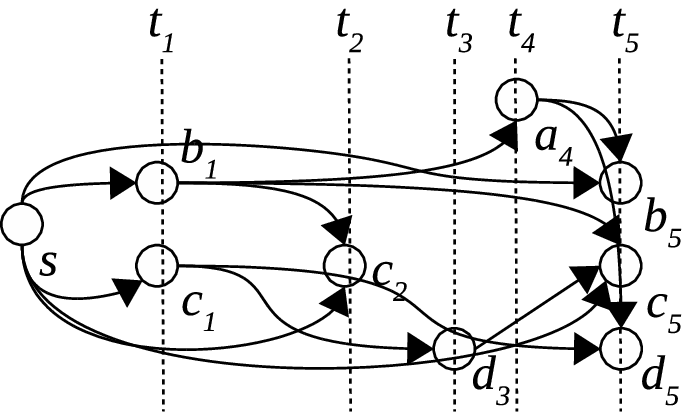}}
\end{center}\caption{An example network $G$ with $V = \{a, b, c, d\}$. In Figure \ref{fig:example_network}, the network communication history is shown in sequence. The link delays are shown on the edges. In Figure \ref{fig:example_history}, the time expanded hypergraph of the network in Figure \ref{fig:example_network} is shown assuming that node $b$ and $c$ start with a message at time $t_1$. In Figure \ref{fig:example_infinite}, we show the corresponding natural information flow graph $G_{\infty}$. In Figure \ref{fig:example_fullmemory} shows the corresponding PNC transform $G_{PNC}$.
}\label{fig:example}\vspace*{-.3cm}
\end{figure*}

\section{Our Results}\label{sec:results}

Given an adversarial schedule and an initial message distribution, the network capacity between the source and any node at any time can be determined. To do so, one enriches the time expanded hypergraph by memory edges, which capture the possibility that nodes \emph{store} knowledge over time. This is achieved by connecting each node $v_t$ in the time expanded hypergraph to its next copy in time $v_{t'}$ with an edge with capacity equal to the amount of information that $v$ can store, i.e., its buffer size $\mu$ (in packets). We assume for simplicity that all nodes have the same amount of memory $\mu$. If all nodes have unlimited buffers, we follow \cite{wu2006trellis} and set $\mu = \infty$. We call this enriched time expanded hypergraph the \emph{(natural) information flow graph} and denote it by $G_{\mu}$. The next lemma confirms the intuition that the information flow graph indeed represents an upper-bound on the amount of information that can be transmitted by \emph{any} algorithm.

\begin{lemma}\label{lemma:mincutbound}
Let $G$ be the time expanded hypergraph for a communication schedule and let $G_{\mu}$ be its natural information flow graph. The the min-cut between the supersource $s$ and a node $v_t$ in $G_{\mu}$ is an upper bound on the amount of information that \emph{any} algorithm can transmit from the sources to node $v$ by time $t$ if all nodes have an active memory of at most $\mu$. 
\end{lemma}

While Lemma~\ref{lemma:mincutbound} provides a simple upper bound on the achievable point-to-point capacity, the more interesting question is whether a given protocol achieves this capacity. While one would hope that the optimality of random linear network coding carries over from the memoryless setting~\cite{ho2006random}, it is not difficult to find protocols that do not achieve this capacity, e.g., the shift-register finite memory network coding protocol in \cite{lun2006analysis}. In the case of the PNC protocol, several results have shown (asymptotic) order optimality in specific stochastic settings \cite{lun2008coding,lun2005further,Lunthesis,wu2006trellis}, or upper bounds on the stopping time in hypergraph theoretic parameters of the topology that are tight up to constant factors in worst-case examples \cite{algebraicgossip-deb-med-choute-06-transinf,mosk2006information,vasudevan2009algebraic,borokhovich2010tight,haeupler2010analyzing,petar}. In the next section, we provide a simpler and a more general approach that proves optimality in \emph{all} the above cases (albeit without providing any bound for concrete stopping times in specific models).

\vspace*{0cm}\subsection{General Approach}

We show that, for many network coding protocols, it is possible to systematically \emph{transform} the time expanded hypergraph into a circuit that \emph{exactly captures} how the protocol uses memory. Given a protocol, a communication schedule, and the corresponding circuit, we prove optimality in three steps. We first show that the circuit indeed simulates the execution of protocol; then apply the results from \cite{ho2006random} for memoryless circuits to show that the protocol achieves the min-cut of this circuit with high probability; and finally show that the min-cut of the circuit is equivalent to the min-cut in $G_\mu$.

To describe our transforms, we note that many network coding protocol proposed so far~\cite{chou2003practical,lun2006analysis,haeupler2011onepacket} are composed of two elementary operations: 1) coding packets together by taking a random linear combination of them, and 2) storing packets in memory. While the coding operation is already naturally captured by the memoryless circuits we show that the storing operation can be simulated by extending a hyperedge (representing a transmission) to all future versions of the recipient(s). Using this observation, we define a hypergraph transformation $(.)_X$ for any given protocol implementation $X$. This transformation takes a time expanded hypergraph $G$ and transforms it to the hypergraph $G_X$ that exactly captures the execution of protocol $X$ on the communication schedule $G$. Note that the hypergraph transformation $(.)_X$ does not just depend on the amount of memory $X$ uses but has to be carefully designed to match the implementation details of protocol $X$.

\vspace*{0cm}\subsection{Protocols and their Transformations}\label{sec:transform}

In this section, we describe the transforms for several protocols. We start with the PNC-protocol from Section \ref{sec:pnc} and then cover two network coding protocols described in \cite{haeupler2011onepacket}: the $\mu$-recombinator and the $\mu$-accumulator protocols. Both protocols are highly efficient variants of PNC, for which any node only stores $\mu$ packets in its buffer. Besides reducing the required memory resources, this also improves the computational cost of network coding, because of the reduced amount of information each coding operation is performed over. The two protocols differ in the way the new set of $\mu$ packets is obtained after a reception of a new packet (and/or generation of a new packet). The $\mu$-recombinator simply picks $\mu$ random packets from the span on the stored packets and the received packets while the more efficient $\mu$-accumulator randomly combines the incoming packet with each of the $\mu$ stored packet individually. The next two definitions present the transformations for the PNC protocol and the $\mu$-recombinator protocol.

\begin{definition}[PNC transform]\label{def:pnctransform}
The PNC-transform $G_{PNC}$ of a time expanded hypergraph $G=(V,V',A)$ is formed by replacing every hyperedge $e \in A$ by it memory closure $\overline{e}$. Here the memory-closure of a hyperedge $e = (v_t, R_e)= (v_t, \{u^1_{t_1}, u^2_{t_2}, ... u^b_{t_b}\}) \in A$ is defined as $\overline{e} = (v_t, \overline{R_e})$ where $\overline{R_e} = \{ u_{t'}\  |  \ \exists u,t : u_t \in R_e \text{ and } t' \geq t\}$. In other words, we extend every hyperedge $e$ to include all future copies of the recipients. 
\end{definition}

\begin{definition}[$\mu$-recombinator transform]\label{def:mutransform}
The $\mu$-re\-com\-bi\-na\-tor transform $G_{\mu\mbox{-recombinator}}$ of a time expanded hypergraph $G=(V,V',A)$ is formed by adding $\mu$ edges from every vertex $v_t \in V'$ to its next copy in time $v_{t'}$ where $t'$ is the smallest $t''>t$ with $v_{t''} \in V'$.
\end{definition}

Note that the two transforms, $G_{PNC}$ and $G_{\mu-\text{recombinator}}$, have an intuitive structure. Extending a hyperedge in $G_{PNC}$ can be interpreted as changing the storage operation of nodes
 to requesting/receiving the exact same packet again whenever the ``stored'' packet is used. For $G_{\mu-\text{recombinator}}$, the $\mu$ memory edges represent that the $\mu$ ``stored'' packets are used to generate the next $\mu$ random packets to be stored. 

\begin{figure}
\begin{center}
\includegraphics[width=0.35\textwidth]{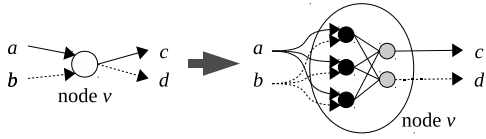}
\end{center}
\caption{Template for $(.)_{\mu\text{-accumulator}}$ with $\mu=3$: The $\mu$ black nodes represent the memory and the gray nodes represent transmissions.}\label{fig:acc1}\vspace*{-.3cm}
\end{figure}

Note that, in general, the network transforms are not necessarily as natural and straight-forward as suggested by Definitions~\ref{def:pnctransform} and \ref{def:mutransform}. One has to be very careful to specify and map all implementation details. Indeed, the transform presented in Definition \ref{def:mutransform} does not exactly capture the protocol described in \cite{haeupler2011onepacket} but instead also recombines its stored packets whenever a packet is send. For simplicity, we consider this variant of the recombinator protocol here. As a final example for a slightly more complicated transformation, we pictorially describe the $\mu$-accumulator transform. We consider the implementation described in \cite{haeupler2011onepacket,lun2006analysis} in which a random multiple of the received packet(s) is added to each stored packet. 
 Its network transform $G_{\mu\text{-accumulator}}$ is formed by first taking the $G_{PNC}$ and then replacing each node according to the template in Figure \ref{fig:acc1}.

\vspace*{0cm}\subsection{Simulation and Optimality Proofs}\label{sec:simulateandoptimality}

Showing that a protocol implementation and its induced hypergraph transformation match is almost always a straight forward inductive proof: 

\begin{theorem}\label{theorem:simulate}
Consider a network using the PNC protocol, and let $G$ be the corresponding time-expanded hypergraph with supersource $s$. Consider the PNC transform $G_{PNC}$ as a circuit as in Section \ref{sec:classic}. If the coding vectors for this circuit are selected independently and uniformly from $F_q$ then this simulates the behavior of the PNC protocol. The message associated with each circuit hyperedge $\overline{e} = (v_t, \overline{R_e})$ in $G_{PNC}$ is the message sent by node $v$ at time $t$. Furthermore, the messages on the incoming hyperedges of $v_t$ in $G_{PNC}$ correspond to the messages stored in memory of node $v$ at time $t$. 
\end{theorem}
\begin{IEEEproof}
For sake of space we present only a proof sketch: In order to prove that the circuit $G_{PNC}$ simulates the execution of the PNC protocol, we need to specify carefully how the randomness used on both sides. For the PNC protocol we assume that a node keeps all received packets (and does not, e.g., keep only innovative packets) and creates any coded packet by drawing random coding coefficient for the packets in the order they were received. We similarly fix the process of choosing the random coding vectors for the circuit to make it match with the PNC protocol. 

Now using an inductive proof over the time (or the topological depth of the nodes in $\overline{G}$), we can show that $G_{PNC}$ simulates the PNC protocol. Firstly, the messages associated with the outgoing hyperedges of the supersource $s$ are by definition the messages generated by the sources. Now consider a node $v$ at time $t$. We assume, without loss of generality, that no node sends a packet when it has not received or generated a message. Thus, $v_t$ has at least one incoming hyperedge from another node $u_{t'}$ where $t' <t$. By construction of $G_{PNC}$, the incoming hyperedges to $v_t$ are from all nodes that have sent a packet to $v$ before time $t$. By induction hypothesis, the incoming hyperedges of $v$ correspond to the messages stored in $v$ in the PNC protocol at time $t$. Since both the circuit $G_{PNC}$ and the PNC protocol linearly combine packets using the same random coefficients, the hypothesis holds for the packets created at node $v$ at time $t$. 
\end{IEEEproof}

\medskip

Given $G_X$ as a representation of the execution of $X$ on the communication schedule $G$ it is easy to state and proof an equivalent of Lemma \ref{lemma:mincutbound}: The amount of source information transmitted from $s$ to $v$ at time $t$ via protocol $X$ is at most the $(s,v_t)$-min-cut in $G_X$. More interestingly, since $G_X$ is memoryless, we can directly apply the results of \cite{ho2006random} to show the converse:

\begin{lemma}\label{lemma:mincutboundprotocol}
Let $G$ be the time expanded hypergraph for a communication schedule and let $G_{X}$ be its transform for the network coding protocol $X$. With probability $1-\eps$, the amount of information transmitted from the sources to node $v$ by time $t$ is exactly the min-cut between the supersource $s$ and a node $v_t$ in $G_{X}$. Here $\eps=O(1/\poly(n))$ is an arbitrarily small inverse polynomial probability given that the coefficient size $\log q$ used in $X$ is $\Theta(\log n)$.
\end{lemma}

All that is left to check is that for the protocols presented here this min-cut is indeed the same as the information theoretical optimum as given by $G_\mu$ in Lemma \ref{lemma:mincutbound}:

\begin{lemma}\label{lemma:equivalent}
Let $G$ be any time expanded hypergraph with supersource $s$. The min-cut between the supersource $s$ and any node $v_t$ is the same in $G_\infty$ and $G_{PNC}$. Furthermore, the same is true for  $G_{\mu}$, $G_{\mu\text{-recombinator}}$, and $G_{\mu\text{-accumulator}}$.
\end{lemma}
\noindent \begin{IEEEproof}
We begin with the equivalence of $G_\infty$ and $G_{PNC}$. For this we transform any integral flow in $G_\infty$ to a valid flow in $G_{PNC}$ and vice versa. Then, we use the min-cut max-flow theorem. The transformation operates on each path in a flow decomposition separately and repeatedly removes flow from $\infty$-edges. Consider a flow-carrying unit-capacity hyperedge $(u_t,w_{t'})$ with an $\infty$-capacity memory-edge $(w_{t'}, w_{t''})$ immediately following it ($t < t' < t''$). We eliminate such $\infty$-edges one-by-one by rerouting the flow directly through $u_t$ to $w_{t''}$ using the extended hyperedges in $G_{PNC}$. This process is flow preserving, respects capacities, and eliminates all $\infty$-edges since every flow path starts with an unit-capacity outgoing hyperedge of $s$. It can be verified that this transformation is also reversible; thus, gives a bijection between integral $(s,v_t)$-flows in $G_\infty$ and integral $(s,v_t)$-flows in $G_{PNC}$. This finishes the proof for $G_{PNC}$.

For $G_{\mu\text{-recombinator}}$, one can use the same strategy, and re-route the flow over the $\mu$-capacity memory-edges in $G_\mu$ to the $\mu$ unit-capacity edges in $G_{\mu\text{-recombinator}}$. 

Similary, for $G_{\mu\text{-accumulator}}$, we first re-route the flow over the $\mu$-capacity memory-edges in $G_\mu$ via the extended hyperedges in $G_{\mu\text{-accumulator}}$ created by the PNC transformation. After the PNC transformation, $G_{\mu\text{-accumulator}}$ is formed by replacing each node according to the template in Figure \ref{fig:acc1}. In $G_{\mu\text{-accumulator}}$, we can re-route the flows of each replaced node since each node $v_t$ in $G_\mu$ carries at most $\mu$-units of flow. This is true by construction: if a node $v$ is receiving at time $t$, node $v_t$ has one out-going memory-edge with capacity $\mu$; if a node $v$ is transmitting at time $t$, then node $v_t$ has only one in-coming memory-edge with capacity $\mu$. 
\end{IEEEproof}

\medskip

Putting everything together finishes our main theorem:

\begin{theorem}\label{theorem:main}
Assume a network and communication model in which the random coding coefficients are independent from the communication schedule. With high probability, the $PNC$, the $\mu$-recombinator, and the $\mu$-accumulator protocols disseminate exactly the maximum amount of information from the sources to every node that \emph{any} protocol using $\mu$ memory could have disseminated. 
\end{theorem}

\section{Conclusion}\label{sec:conclusions}
In this paper, we resolve the question of optimality for the well-studied PNC protocol and similar network coding protocols. These protocols use the memory of nodes to produce coded packets that reflect everything received so far. We show that an implementation of such a protocol induces a transformation that maps any execution to an instance of the classical memoryless setting. This shows that PNC solves the many-to-all multicast problem in \emph{any} non-adaptive dynamic network model in \emph{optimal} time. Differently phrased, PNC makes on-the-fly the optimal decision of what information a node should send out without knowing anything about the network topology or even which other node will receive this information. 

Even more interestingly, this remains true if one restricts the nodes to use limited size buffers. We show that both the $\mu$-recombinator and the $\mu$-accumulator protocol\cite{haeupler2011onepacket} achieve optimal performance in this setting: with high probability they stop exactly within the time in which in hindsight it was possible to route packets given the buffer constraint, i.e., given that the buffer at each node never exceeds the limit. Alternatively, one can interpret this result as PNC making on-the-fly optimal decisions on which information to keep in the limited memory. This shows that, even without any feedback\cite{sundararajan2008arq} or complicated explicit memory management, these PNC variants preserve the capacity achieving performance of PNC as long as minimal buffer sizes are available. 

This paper also implies that determining stopping times for PNC is equivalent to determining the connectivity of a network or, more generally, to determining the network capacity. For many settings, obtaining good bounds or characterizations for the network capacity remains an interesting open question. Recently, significant progress was made in this direction for both the PNC protocol~\cite{haeupler2010analyzing} and its finite memory variants~\cite{haeupler2011onepacket}. We are hopeful that the insights provided here will be helpful in further advances.



\begin{thebibliography}{10}
\providecommand{\url}[1]{#1}
\csname url@samestyle\endcsname
\providecommand{\newblock}{\relax}
\providecommand{\bibinfo}[2]{#2}
\providecommand{\BIBentrySTDinterwordspacing}{\spaceskip=0pt\relax}
\providecommand{\BIBentryALTinterwordstretchfactor}{4}
\providecommand{\BIBentryALTinterwordspacing}{\spaceskip=\fontdimen2\font plus
\BIBentryALTinterwordstretchfactor\fontdimen3\font minus
  \fontdimen4\font\relax}
\providecommand{\BIBforeignlanguage}[2]{{%
\expandafter\ifx\csname l@#1\endcsname\relax
\typeout{** WARNING: IEEEtran.bst: No hyphenation pattern has been}%
\typeout{** loaded for the language `#1'. Using the pattern for}%
\typeout{** the default language instead.}%
\else
\language=\csname l@#1\endcsname
\fi
#2}}
\providecommand{\BIBdecl}{\relax}
\BIBdecl

\bibitem{li2003linear}
S.~Li, R.~Yeung, and N.~Cai, ``{Linear network coding},'' \emph{Transactions on
  Information Theory (TransInf)}, vol.~49, no.~2, pp. 371--381, 2003.

\bibitem{ho2006random}
T.~Ho, M.~M{\'e}dard, R.~Koetter, D.~Karger, M.~Effros, J.~Shi, and B.~Leong,
  ``{A random linear network coding approach to multicast},''
  \emph{Transactions on Information Theory (TransInf)}, vol.~52, no.~10, pp.
  4413--4430, 2006.

\bibitem{koetter2003algebraic}
R.~Koetter and M.~M\'edard, ``{An algebraic approach to network coding},''
  \emph{Transactions on Networking (TON)}, vol.~11, no.~5, pp. 782--795, 2003.

\bibitem{chou2003practical}
P.~Chou, Y.~Wu, and K.~Jain, ``{Practical network coding},'' in \emph{Proc. of
  the 41st Allerton Conference on Communication Control and Computing},
  vol.~41, 2003, pp. 40--49.

\bibitem{lun2008coding}
D.~Lun, M.~M{\'e}dard, R.~Koetter, and M.~Effros, ``{On coding for reliable
  communication over packet networks},'' \emph{Physical Communication}, vol.~1,
  no.~1, pp. 3--20, 2008.

\bibitem{lun2005further}
------, ``{Further results on coding for reliable communication over packet
  networks},'' in \emph{Proc. of the International Symposium on Information
  Theory (ISIT)}, 2005, pp. 1848--1852.

\bibitem{Lunthesis}
D.~Lun, ``Efficient operation of coded packet networks,'' Ph.D. dissertation,
  Massachusetts Institute of Technology, 2006.

\bibitem{wu2006trellis}
Y.~Wu, ``{A trellis connectivity analysis of random linear network coding with
  buffering},'' in \emph{Proc. of the International Symposium on Information
  Theory (ISIT)}, 2006, pp. 768--772.

\bibitem{algebraicgossip-deb-med-choute-06-transinf}
S.~Deb, M.~M\'edard, and C.~Choute, ``Algebraic gossip: a network coding
  approach to optimal multiple rumor mongering,'' \emph{Transactions on
  Information Theory (TransInf)}, vol.~52, no.~6, pp. 2486 -- 2507, 2006.

\bibitem{mosk2006information}
D.~Mosk-Aoyama and D.~Shah, ``{Information dissemination via network coding},''
  in \emph{Proc. of the International Symposium on Information Theory (ISIT)},
  2006, pp. 1748--1752.

\bibitem{vasudevan2009algebraic}
D.~Vasudevan and S.~Kudekar, ``{Algebraic gossip on Arbitrary Networks},''
  \emph{ArXiv:0901.1444}, 2009.

\bibitem{borokhovich2010tight}
M.~Borokhovich, C.~Avin, and Z.~Lotker, ``{Tight bounds for algebraic gossip on
  graphs},'' in \emph{Proc. of the International Symposium on Information
  Theory (ISIT)}, 2010, pp. 1758--1762.

\bibitem{haeupler2010analyzing}
B.~Haeupler, ``{Analyzing Network Coding Gossip Made Easy},'' \emph{Proc. of
  the 43nd Symposium on Theory of Computing (STOC)}, 2011.

\bibitem{petar}
P.~Maymounkov, N.~Harvey, and D.~Lun, ``{Methods for efficient network
  coding},'' in \emph{Proc. of the 44th Allerton Conference on Communication,
  Control, and Computing}, 2006.

\bibitem{haeupler2011onepacket}
B.~Haeupler and M.~M{\'e}dard, ``{One Packet Suffices - Highly Efficient
  Packetized Network Coding With Finite Memory},'' \emph{ArXiv}, 2011.

\bibitem{lun2006analysis}
D.~S. Lun, P.~Pakzad, C.~Fragouli, M.~M\'edard, and R.~Koetter, ``{An analysis
  of finite-memory random linear coding on packet streams},'' in \emph{Proc. of
  the International Symposium on Modeling and Optimization in Mobile, Ad Hoc
  and Wireless Networks (WiOpt)}, 2006, pp. 1--6.

\bibitem{sundararajan2008arq}
K.~Sundararajan, D.~Shah, and M.~M{\'e}dard, ``{ARQ for network coding},'' in
  \emph{Proc. of the International Symposium on Information Theory (ISIT)},
  2008, pp. 1651--1655.

\end{thebibliography}
\end{document}